\RequirePackage{fix-cm}
\documentclass[twocolumn,epjc3]{svjour3}  
\smartqed  % flush right qed marks, e.g. at end of proof
\RequirePackage{graphicx}
\RequirePackage{mathptmx}      % use Times fonts if available on your TeX system
\RequirePackage{latexsym}
\RequirePackage[numbers,sort&compress]{natbib}
\RequirePackage[colorlinks,citecolor=blue,urlcolor=blue,linkcolor=blue]{hyperref}
\usepackage{lineno}
\usepackage{isotope}
\usepackage{subcaption}
\usepackage{mathtools} % text above rightarrow
\usepackage{braket}
\usepackage{comment}
\journalname{Eur. Phys. J. A}

\def\makeheadbox{\relax} %uncomment for showing journal

\begin{document}
	% \linenumbers
	\title{Improved FIFRELIN de-excitation model for neutrino applications%\thanksref{t1}
	}
	% \title{Improvements on the FIFRELIN de-excitation model for the isotopes $\mathbf{^{156,158}}$Gd%\thanksref{t1}
	
	
	\newcommand{\MPIK}{\affiliation{Max-Planck-Institut f\"ur Kernphysik, Saupfercheckweg 1, 69117 Heidelberg, Germany}}
	\newcommand{\LAPP}{\affiliation{Univ. Savoie Mont Blanc, CNRS, LAPP-IN2P3, 74000 Annecy, France}}
	\newcommand{\LPSC}{\affiliation{Univ.~Grenoble Alpes, CNRS, Grenoble INP, LPSC-IN2P3, 38000 Grenoble, France}}
	\newcommand{\CEA}{\affiliation{IRFU, CEA, Universit\'e Paris-Saclay, 91191 Gif-sur-Yvette, France}}
	\newcommand{\ILL}{\affiliation{Institut Laue-Langevin, CS 20156, 38042 Grenoble Cedex 9, France}}
	
	\author{H. Almaz\'an\thanksref{e1,MPIK}
		\and 
		L. Bernard\thanksref{e2,LPSC}
		\and
		A. Blanchet\thanksref{e3,IRFU}
		\and
		A. Bonhomme\thanksref{MPIK,IRFU}
		\and
		C. Buck\thanksref{MPIK}
		\and
		A. Chalil\thanksref{e4,IRFU}
		\and
		A. Chebboubi\thanksref{CADARACHE}
		\and
		P. del Amo Sanchez\thanksref{LAPP}
		\and
		I. El Atmani\thanksref{e5,IRFU}
		\and
		L. Labit\thanksref{LAPP}
		\and
		J. Lamblin\thanksref{LPSC}
		\and
		A. Letourneau\thanksref{IRFU}
		\and
		D. Lhuillier\thanksref{IRFU}
		\and
		M. Licciardi\thanksref{LPSC}
		\and
		M. Lindner\thanksref{MPIK}
		\and
		O. Litaize\thanksref{CADARACHE}
		\and
		T. Materna\thanksref{IRFU}
		\and
		H. Pessard\thanksref{LAPP}
		\and
		J.-S. R\'eal\thanksref{LPSC}
		\and
		J.-S. Ricol\thanksref{LPSC}
		\and
		C. Roca\thanksref{MPIK}
		\and 
		R. Rogly\thanksref{IRFU}
		\and
		T. Salagnac\thanksref{e6,LPSC}
		\and
		V. Savu\thanksref{IRFU}
		\and
		S. Schoppmann\thanksref{e7,MPIK}
		\and
		T. Soldner\thanksref{ILL}
		\and
		A. Stutz\thanksref{LPSC}
		\and
		L. Thulliez\thanksref{IRFU}
		\and
		M. Vialat\thanksref{ILL}
	}

	\thankstext{e1}{Present address: Donostia International Physics Center, Paseo Manuel Lardizabal, 4, 20018 Donostia-San Sebastian, Spain}
	\thankstext{e2}{Present address: Ecole Polytechnique, CNRS/IN2P3, Laboratoire  Leprince-Ringuet, 91128 Palaiseau, France}
	\thankstext{e3}{Present address: LPNHE, Sorbonne Universit\'e, Universit\'e de Paris, CNRS/IN2P3, 75005 Paris, France}
	\thankstext{e4}{Corresponding author: achment.chalil@cea.fr}
	\thankstext{e5}{Present address: Hassan II University, Faculty of Sciences, A\"in Chock, BP 5366 Maarif, Casablanca 20100, Morocco}
	\thankstext{e6}{Present address: Institut de Physique Nucl\'eaire de Lyon, CNRS/IN2P3, Univ. Lyon, Universit\'e Lyon 1, 69622 Villeurbanne, France}
	\thankstext{e7}{Present address: University of California, Department of Physics, Berkeley, CA 94720-7300, USA and Lawrence Berkeley National Laboratory, Berkeley, CA 94720-8153, USA}
	
	\institute{{Max-Planck-Institut f\"ur Kernphysik, Saupfercheckweg 1, 69117 Heidelberg, Germany}\label{MPIK}
		\and
		{Univ.~Grenoble Alpes, CNRS, Grenoble INP, LPSC-IN2P3, 38000 Grenoble, France}\label{LPSC}
		\and
		IRFU, CEA, Universit\'{e} Paris-Saclay, 91191 Gif-sur-Yvette, France\label{IRFU}
		\and
		CEA, DES, IRESNE, DER, Cadarache, 13108 Saint-Paul-Lez-Durance, France \label{CADARACHE}
		\and 
		Univ. Savoie Mont Blanc, CNRS, LAPP-IN2P3, 74000 Annecy, France \label{LAPP}
		\and
		Institut Laue-Langevin, CS 20156, 38042 Grenoble Cedex 9, France \label{ILL}
	}
	
	\date{Received: date / Accepted: date}

	\maketitle
	
	\begin{abstract}
		
		The precise modeling of the de-excitation of Gd isotopes is of great interest for experimental studies of neutrinos using Gd-loaded organic liquid scintillators.
		The FIFRELIN code was recently used within the purposes of the STEREO experiment for the modeling of the Gd de-excitation after neutron capture in order to achieve a good control of the detection efficiency. In this work, we report on the recent additions in the FIFRELIN de-excitation model with the purpose of enhancing further the de-excitation description. Experimental transition intensities from EGAF database are now included in the FIFRELIN cascades, in order to improve the description of the higher energy part of the spectrum. Furthermore, the angular correlations between $\gamma$ rays are now implemented in FIFRELIN, to account for the relative anisotropies between them.  In addition, conversion electrons are now  treated more precisely in the whole spectrum range, while the subsequent emission of X rays is also accounted for. The impact of the aforementioned improvements in FIFRELIN is tested by simulating neutron captures in various positions inside the STEREO detector. A repository of up-to-date FIFRELIN simulations of the Gd isotopes is made available for the community, with the possibility of expanding for other isotopes which can be suitable for different applications.

		%  \PACS{ 23.20.En \and 21.10.Hw \and 07.05.Tp}
	\end{abstract}

	\section{Introduction}
	\label{intro}
	
	FIFRELIN~\cite{LITAIZE201251,Litaize2015,REGNIER201619} is a Monte Carlo code developed to model the fission process for reactor applications. FIFRELIN has the capability to be run in a decay mode, allowing the modeling of the de-excitation of any isotope. Recently, simulated cascades for the isotopes \isotope[156,158]{Gd} have been used in the STEREO experiment, yielding an improved agreement with the data~\cite{Improved_stereo_Almazan2019}. These cascades have been made publicly available for use in other suitable applications~\cite{zenodo}. Moreover, within the CRAB method, which was recently proposed for the calibration of bolometers in the 100 eV region~\cite{Thulliez_2021}, the FIFRELIN cascades of W isotopes were used for its feasibility study.
	
	The wide use of Gd-loaded organic liquid scintillators require a precise knowledge of the nuclear de-excitation of Gd. Such detectors are used for the measurement of the antineutrino flux coming from nuclear reactors. Experiments such as STEREO~\cite{Allemandou_2018_The_STEREO_EXP, STEREO_improved_sn_constraints_PhysRevD.102.052002, Almazan_JPhysG2021}, Daya Bay~\cite{Daya_bay_AN2016133,Daya_Bay_PhysRevLett.123.111801} and RENO~\cite{RENO_PhysRevLett.125.191801} have employed Gd-loaded liquid scintillators for their measurements.  In these cases, the detection of an electron antineutrino $\overline\nu_e$ is registered through the Inverse-Beta-Decay process (IBD) on the protons of the liquid:
	\begin{equation}
	\overline{\nu}_{e} + p \rightarrow e^+ + n
	\end{equation}

	Despite the improved description of FIFRELIN when compared to data from the STEREO experiment~\cite{Improved_stereo_Almazan2019}, there were still missing aspects of the de-excitation process that could be added to further improve the overall description of the de-exciting Gd nuclei. Examples of such aspects are the anisotropic emission of the $\gamma$ rays with respect to the previously emitted ones and the X ray emission after Internal Conversion (IC). Furthermore, the reliance on theoretical models for the primary $\gamma$ rays could be further improved with the addition of evaluated transition intensities. Moreover, the version of FIFRELIN used in~\cite{Improved_stereo_Almazan2019} uses an outdated version of RIPL-3 database (RIPL-3 v.2015).

	In this work, we describe the latest improvements to the FIFRELIN de-excitation model. An updated version of RIPL-3 database (RIPL-3 v.2020)~\cite{CAPOTE20093107} is now used as input for the present FIFRELIN cascades. In addition, the EGAF~\cite{EGAF} database has been implemented in FIFRELIN, to account for the primary $\gamma$ ray intensities, i.e. the transitions that depopulate the neutron-separation energy level, for the isotopes \isotope[156,158]{Gd}. This is expected to improve the de-excitation description of primary $\gamma$ rays, as only theoretical models were used in the previous case~\cite{Improved_stereo_Almazan2019}.  Furthermore, the physics of $\gamma$ directional correlations has been implemented in order to account for the anisotropic emission of the $\gamma$ rays with respect to the previously emitted ones. Internal Conversion (IC) and Internal Pair Conversion (IPC) coefficients are updated up to 6 MeV, using the BrIcc code~\cite{KIBEDI2008202} and by taking into account the electronic binding energies. The subsequent X ray emission after IC is also accounted for in the new versions of the cascades. Furthermore, during the IPC process, a positron is now emitted along with the electron, in opposite directions.
	
	All the aforementioned improvements are now available in an updated repository of Gd cascades, to be used for any suitable application~\cite{zenodo_2022}, with the potential of expanding the repository in order to include cascades for other nuclei that are used in similar applications.

	%---------------------------------------------------------------------------
	\section{The FIFRELIN de-excitation model}
	\label{sec: fifrelin}
	
	The FIFRELIN Monte Carlo code builds the low-energy level scheme with extensive use of known evaluated nuclear levels from the RIPL-3 database. RIPL-3 contains the necessary input parameters for nuclear reactions and nuclear data evaluations~\cite{CAPOTE20093107}. For this work, the updated version RIPL-3 v.2020 has been used.
	FIFRELIN builds the de-excitation level scheme of a nucleus for three different regions: for $(E<E_{RIPL})$, where $E_{RIPL}$ corresponds to the level below of which the level scheme is considered completely known, the level scheme is constructed only from the RIPL-3 database \cite{CAPOTE20093107}. In the intermediate energy range, $E_{RIPL}<E<E_{limit}$, a combination between levels from RIPL-3 and theoretical levels sampled from spin-dependent level density models are used. Here, $E_{limit}$ corresponds to a level density of $5 \times 10^4$ levels/MeV.
	
	In the high energy interval, $E_{limit}<E<S_n$, where $S_n$ is the neutron-separation energy, FIFRELIN samples the energy levels exclusively from level density models. The updated values of $E_{RIPL}$, $E_{limit}$ and $S_n$ for the isotopes \isotope[156,158]{Gd} are tabulated in Table~\ref{tab: energy values ripl}. The reason for the addition of theoretical models is that the complete level scheme of a nucleus is not known. This holds especially for the continuous part of the spectrum which corresponds to transitions from levels near the neutron separation energy $S_n$. 
	
	In the higher energy region, $E_{limit}<E<S_n$, the energy spectrum is considered continuous and the energy levels are described in bins with a bin width of 1 keV, in order to reduce computation time. The initial and final energies of a transition are sampled inside the bin. Each level is assigned a spin and parity given by a model.

	\begin{table}[t]
		\centering
		\caption{Updated table of critical energy values (in MeV) for the \isotope[156,158]{Gd}, as used in FIFRELIN.}
		\resizebox{0.4\textwidth}{!}{ 
			\begin{tabular}{c|c|c|c|c}
				
				Isotope & $E_{RIPL}$ & $E_{limit}$ & $S_n$ & $E_M$ \\ \hline\hline
				\isotope[156]{Gd} & 1.366 & 5.117 & 8.536 & 6.439 \\
				\isotope[158]{Gd} & 1.481 & 5.191 & 7.937 & 6.633 
			\end{tabular}
		}
		
		\label{tab: energy values ripl}
	\end{table}

	The nuclear level density model used is given by the relation:
	\begin{equation}
	\rho(E,J,\pi) = \rho_{tot}(E) P(J|E) P(\pi),
	\end{equation}
	where $\rho_{tot}(E)$ is the total nuclear level density, which corresponds to the composite Gilbert-Cameron Model~\cite{gilbert_1965_09000a}:
	\begin{equation}
	\rho^{CGCM}_{tot} (E) = \left\{
	\begin{array}{ll}
	\rho^{CTM}_{tot} (E)  & \mbox{if } E \leq E_M \\
	\rho^{FGM}_{tot} (E)  & \mbox{if } E > E_M
	\end{array}
	\right.
	\end{equation}
	Here, $\rho^{CTM}_{tot}$ is the level density as predicted by the Constant Temperature Model (CTM)~\cite{Rosenzweig_PhysRev.105.950} and $\rho^{FGM}_{tot}$ is the level density as predicted by the Fermi Gas Model. The value $E_M$ is the matching energy, and is defined as the energy where the two models have equal values of level density and its first derivative to ensure continuity. A complete description of the models can be found in~\cite{CAPOTE20093107}.
	
	The spin distribution function $P(J|E)$ is given by the relation:
	\begin{equation}
	P(J|E) = \frac{2J+1}{2 \sigma^2(E)} \exp \left( - \frac{(J+1/2)^2}{2 \sigma^2(E)} \right),
	\end{equation}\label{eq: spin distribution function}
	where $\sigma(E)$ is the spin-cutoff parameter~\cite{CAPOTE20093107,Bethe_1936_PhysRev.50.332}. The parity distribution $P(\pi)$ is taken as $0.5$ for both parity signs, meaning that a level with an unknown parity has equal probability of having a positive or a negative parity value.
	
	Once the complete level scheme of the nucleus is generated, the code samples one transition from all possibilities, depending on the values of the emission probabilities. For higher energies near the continuum a statistical treatment is necessary for $\gamma$ de-excitation with the use of $\gamma$-ray strength function models.  
	
	The statistical treatment of an excited nucleus near continuum requires the use of the average transmission coefficient $T_\gamma$, which is given by the relation:
	\begin{equation}
	T_\gamma(E_\gamma,XL) = E_\gamma^{2L+1} f_{XL}(E_\gamma) / \rho (E_i,J_i^\pi)
	\end{equation}
	where $E_\gamma$ is the transition energy depopulating the level with energy $E_i$ and spin-parity $J_i^\pi$, $X$ is the type ($X=E$ for electric and $X=M$ for magnetic transitions), $L$ the multipolarity of the transition and $f_{XL}$ is the $\gamma$-strength function. In this work, FIFRELIN uses two models for the $\gamma$-strength function depending on their multipolarity. For electic dipole  transitions, the $\gamma$-strength function is given by the Enhanced Generalized Lorentzian model (EGLO)~\cite{Kopecky_Uhl_1990_PhysRevC.41.1941}. For other types of multipoles, the Standard Lorentzian model (SLO) is used \cite{BRINK1957215}.

	\begin{figure}[t]
		\centering
		\includegraphics[width=0.45\textwidth]{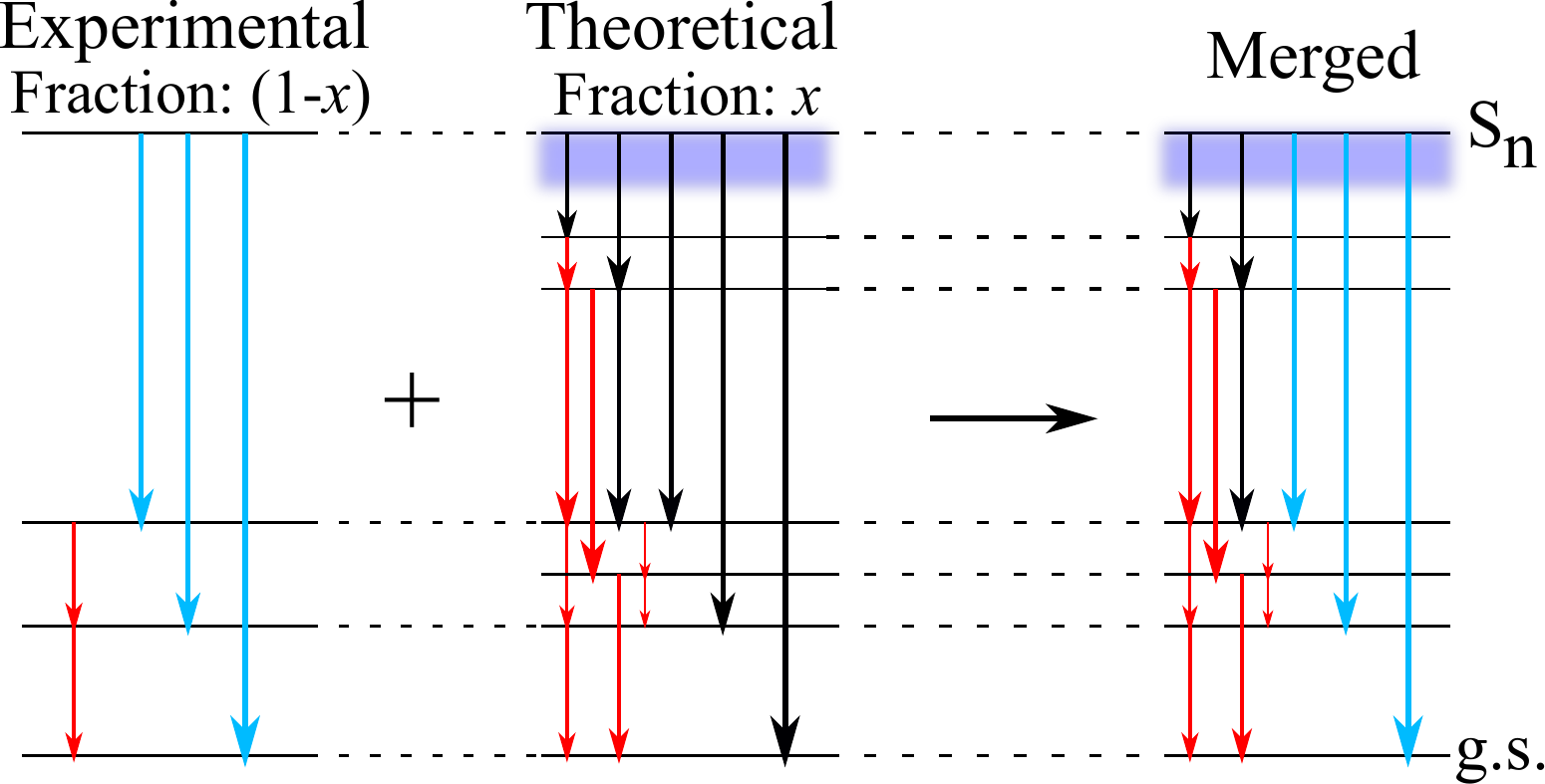}
		\caption{Procedure to simulate the level scheme using both RIPL3 and EGAF databases. A purely ``experimental" simulation using the primary transitions from EGAF (in cyan) is combined with a simulation using RIPL-3 (red) and theoretical models (black). The weighted sum of the two simulations give the final "merged" simulation. Primary transitions that are present in both the experimental and theoretical simulations are removed from the theoretical part, to avoid double counting. See text for details.}
		\label{fig: schema de calcul}
	\end{figure}

	\begin{figure}[t]
		\centering
		\begin{subfigure}[c]{0.45\textwidth}
			\includegraphics[width=\textwidth]{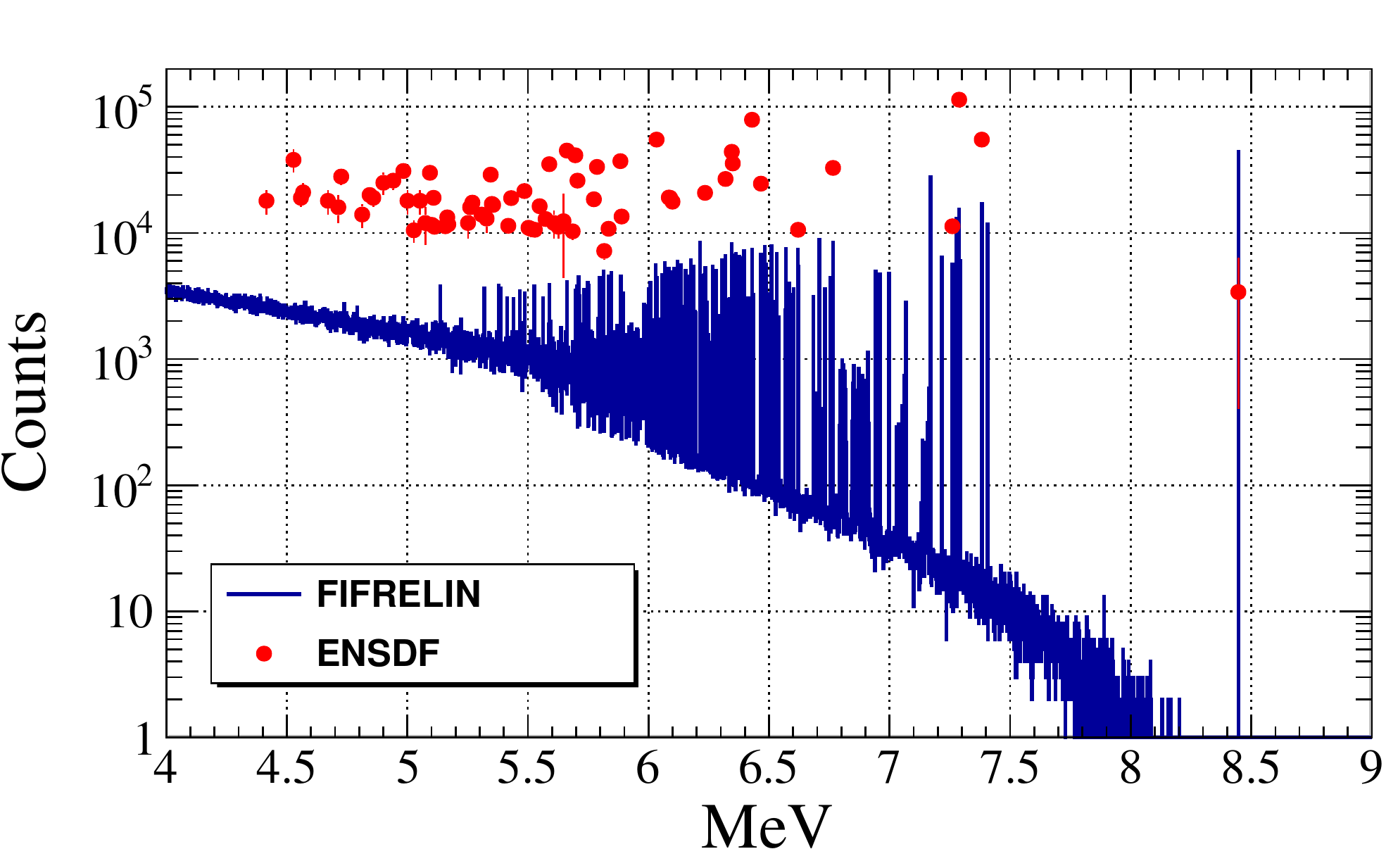}
			\centering
			\caption{}
			\label{fig: 156 theorik}
		\end{subfigure}
		\begin{subfigure}[c]{0.45\textwidth}
			\includegraphics[width=\textwidth]{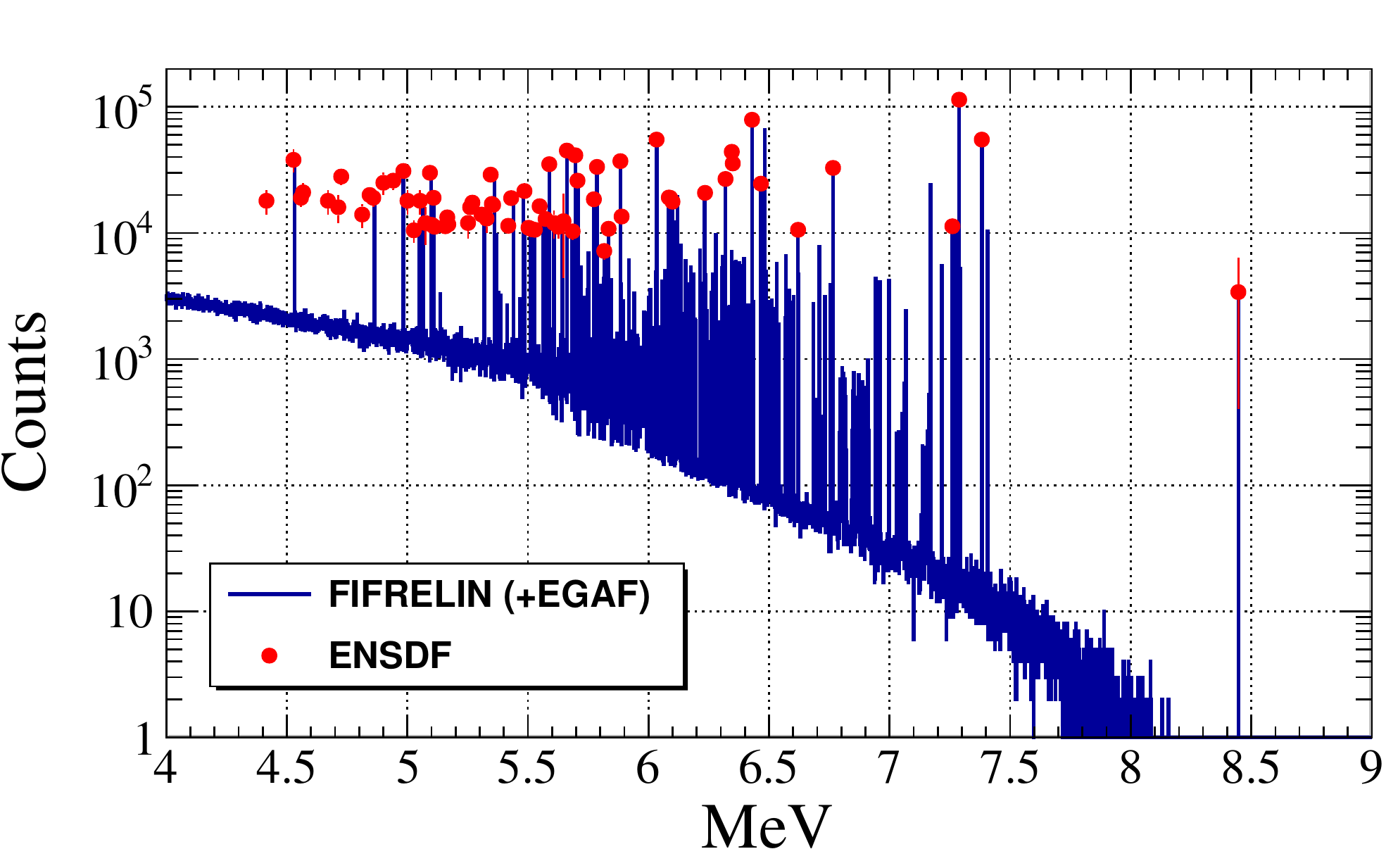}
			\centering
			\caption{}
			\label{fig: 156 egaf}
		\end{subfigure}
		\caption{(a) Higher energy $\gamma$ spectrum of \isotope[156]{Gd} with FIFRELIN, using only theoretical models (b) The same but with FIFRELIN using a combination of EGAF transition intensities and theoretical models. }
		\label{fig: EGAF}
	\end{figure}

	%-----------------------------------------------------------------------------
	
	\section{Upgrade of the de-excitation process in FIFRELIN}
	\label{sec: fifrelin_addition}
	
	\subsection{Implementation of EGAF database}
	
	The Evaluated Gamma Activation File (EGAF)~\cite{Firestone_2006_EGAF_doi:10.1063/1.2187849,EGAF} is a database of prompt and delayed neutron-capture  $\gamma$-ray cross sections. The database consists of data acquired from measurements performed at the Budapest Research Reactor in combination with data from literature~\cite{Firestone_2006_EGAF_doi:10.1063/1.2187849}. The measurements were performed on natural elemental targets. Additional $\gamma$ rays were placed into the Budapest dataset by comparison with expected transitions from the Table of Isotopes \cite{table_of_isotopes}.

	The implementation of the EGAF database in FIFRELIN accounts for the primary $\gamma$ rays emitted from the initial excited state. In the case of thermal-neutron capture, this level corresponds to the neutron-separation energy $S_n$. The procedure to simulate the level scheme is illustrated in Fig.~\ref{fig: schema de calcul}. Firstly, an "experimental" simulation is performed with FIFRELIN, using only levels from the EGAF database. No theoretical models are used in this simulation. Then, a "theoretical" simulation using the standard procedure of FIFRELIN is simulated, using the RIPL-3 database along with the theoretical models described in Section~\ref{sec: fifrelin}. In order to merge the simulation, a weighting factor $x$ has to be estimated, which corresponds to the percentage of the primary transitions of EGAF. Then, the merged simulation is constructed by using a percentage of $x$ cascades from the EGAF simulation and $1-x$ cascades from the theoretical one. For \isotope[156]{Gd}, $x=12.64\%$ while for \isotope[158]{Gd}, $x=18.16\%$.

	The combination of these two databases has yielded a much better description in the higher energy part of the gamma spectrum. Fig~\ref{fig: EGAF} shows that the inclusion of the EGAF data in the FIFRELIN simulation brings the description of all discrete levels a lot closer to the ENSDF~\cite{NNDC} values, which are taken as a reference. Work is in progress for an automated interface with ENSDF that would allow an easier use of FIFRELIN for any target isotope.

	\subsection{$\gamma$-directional correlations}
	
	In parallel to the improved set of discrete levels taken from the nuclear databases, the treatment of the direction of emission of the $\gamma$ rays has been refined. The formal theory of angular correlations has been used, based on the statistical tensor formalism~\cite{Rose_Brink_Revmodphys_1967,Hamilton1975_electromagnetic}. Their calculation is essential for the determination of the probability distribution functions which describe the directions of $\gamma$ rays in the cascade. For a cascade of $\gamma$ rays starting from an initial state $J_0$ and ending to a state $J_n$:
	\begin{equation}\label{eq:cascade1}
	J_0 \xrightarrow[]{\gamma_0} J_1 \xrightarrow[]{\gamma_1} ...  \xrightarrow[]{\gamma_{n-1}} J_n
	\end{equation}
	a set of statistical tensors can be calculated, containing the information on the orientation of the initial state $J_0$. Then, the probability distribution functions for each $\gamma$ can be evaluated and used for the generation of directions $(\theta_i,\phi_i)$ of the $i$-th $\gamma$ ray. The implementation of the angular correlations on FIFRELIN, along with its theoretical description, is described in detail in~\cite{Chalil2022}.

	\subsection{Treatment of conversion electrons and X rays}
	
	IC and IPC coefficients are accounted for using BrIcc code based on the Dirac-Fock calculations under the "Frozen Orbital" approximation~\cite{KIBEDI2008202} within the energy interval from ($E_{shell}+1$) keV up to 6 MeV, where $E_{shell}$ is the electron shell energy. In the previous version of FIFRELIN cascades~\cite{Improved_stereo_Almazan2019}, the kinetic energy of the conversion electron was assigned the energy of the corresponding transition. In the present version, the process is treated more accurately by accounting the electronic binding energies. For the case of IPC, a positron is now emitted together with the electron, in opposite directions.
	
	In addition, a treatment of X ray emission has been also implemented in the newer FIFRELIN cascades. When a conversion electron is emitted, the remaining residual energy is then used to emit an X ray. In this way, the sum of the total energy of the cascade is always equal to the initial energy of the nucleus $S_n$.
	
	%---------------------------------------------------------------------------
	\section{Application to the STEREO detector}
	\label{sec:Application STEREO}
	
\begin{figure}[t]
		\centering
		\includegraphics[width=0.45\textwidth]{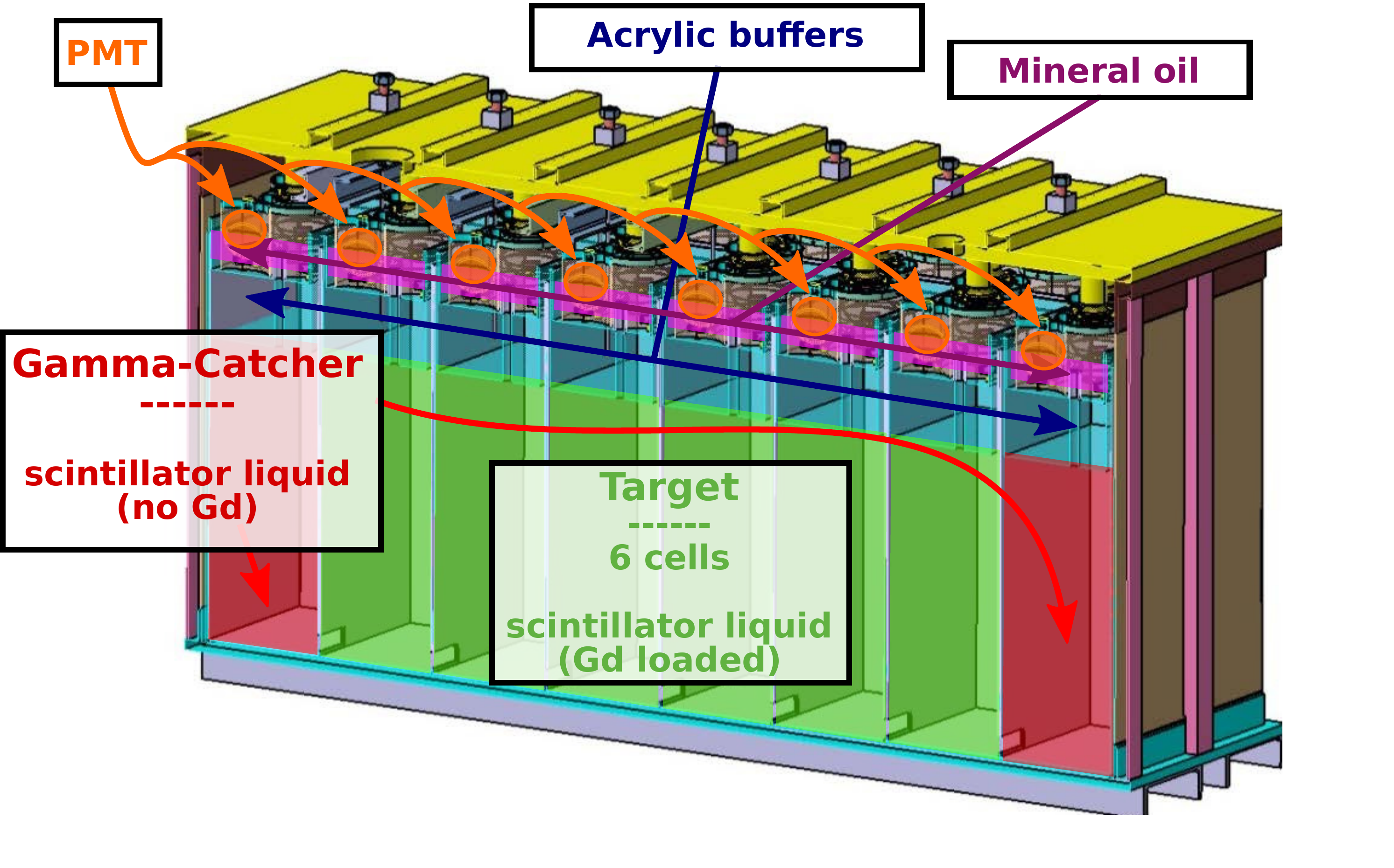}
		\centering
		\label{fig: 156 theorik}
		\caption{Layout of the STEREO detector. The six central Gd-loaded target cells constitute the neutrino target, and are surrounded by a crown of unloaded liquid to mitigate gamma leakages out of the target volume and impact of external background.}
		\label{fig: STEREO detector}
	\end{figure}

	The final test of all these new ingredients of FIFRELIN is the comparison between the simulated and measured spectra after neutron-capture inside a Gd-loaded scintillator. For this purpose, data taken from the STEREO detector are compared with simulations using the updated FIFRELIN cascades. This is an important aspect for all neutrino detection experiments by the IBD process since the accuracy of the detection efficiency directly depends on the control of these spectra.  The layout of STEREO detector is shown on Fig~\ref{fig: STEREO detector}. The detector is composed of six cells of Gd-loaded liquid scintillator (target), surrounded by four cells of Gd-free scintillator (gamma catcher). More details about the STEREO detector can be found in~\cite{Allemandou_2018_The_STEREO_EXP}.
	
	In the STEREO experiment the neutron response is monitored with regular calibration runs where an americium-beryllium (AmBe) source is deployed in 5 of the 6 cells, successively at 5 different heights (10, 30, 45, 60 and 80 cm from the bottom). The neutrons are produced through a two-step process: an Am decay first emits an $\alpha$ particle, which then interacts with the a Be nucleus $\alpha+^9$Be $\rightarrow ^{12}$C + n. In about 60\% of the cases the \isotope[12]{C} isotope is produced in an excited state decaying with a 4.4 MeV $\gamma$ ray. The prompt signal of this source is thus the sum of the energy deposits from proton recoils induced by the few MeV neutron and the high energy $\gamma$ ray. The neutron capture signal is selected by requesting a second energy deposit in a 100 $\mu s$ time window following a first large energy deposit (between 4 and 7 MeV). The size of this time window is set according to the 16 $\mu s$ capture time of a neutron in the target volume of the STEREO detector. Because of the high activity of the source ($\sim 15 \times 10^3$ n/s) special care is taken to the statistical subtraction of accidental pairs of events. The complete description of the statistical analysis of the STEREO data is beyond the scope of present work, and will be thoroughly presented in an upcoming publication of the STEREO collaboration.

	% %

	\begin{figure*}[ht!]
		\centering
		
		\begin{subfigure}[c]{0.47\textwidth}
			\includegraphics[width=\textwidth]{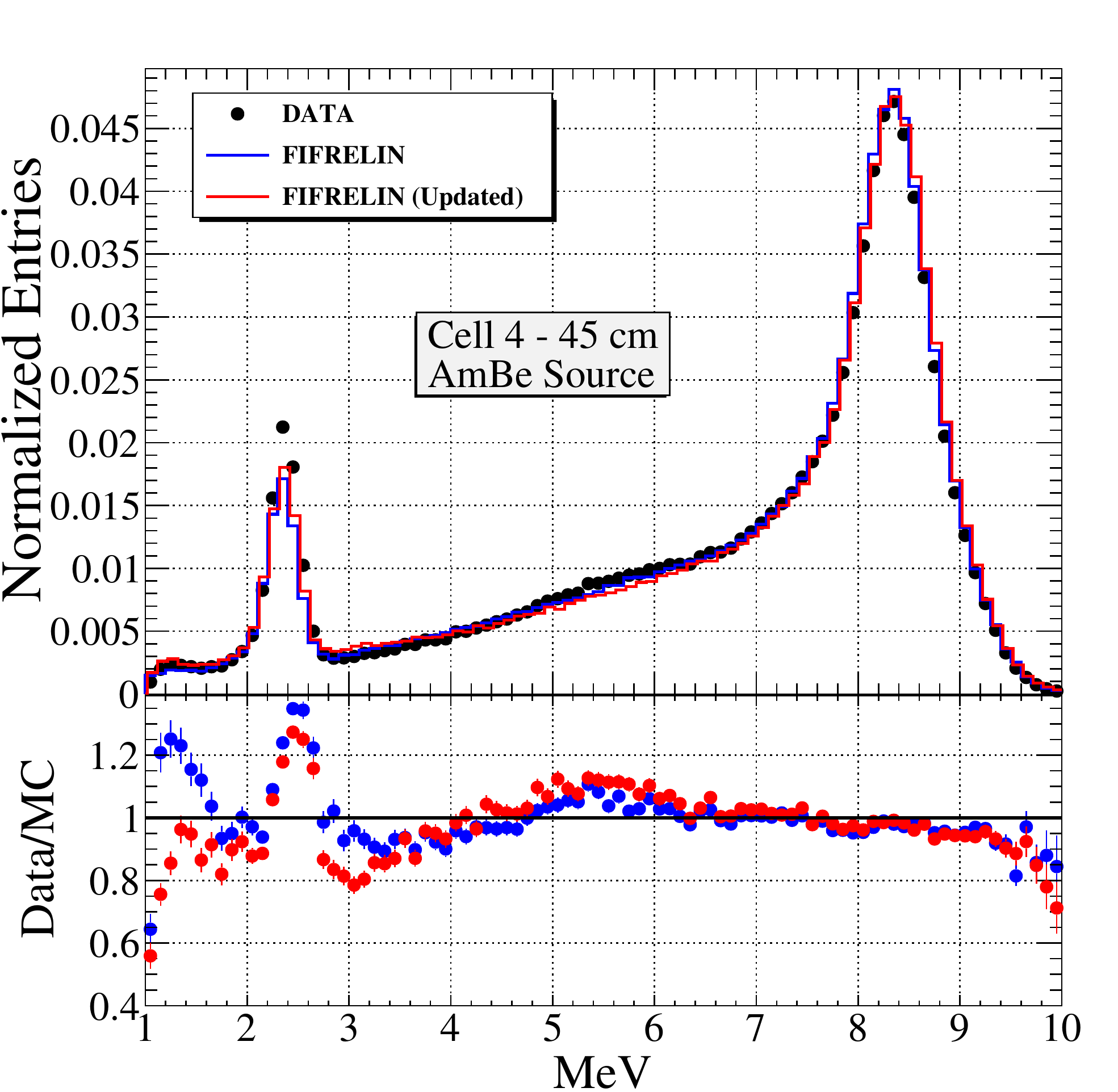}
			\centering
			\caption{}
			\label{fig: cell center}
		\end{subfigure}%
		\begin{subfigure}[c]{0.47\textwidth}
			\includegraphics[width=\textwidth]{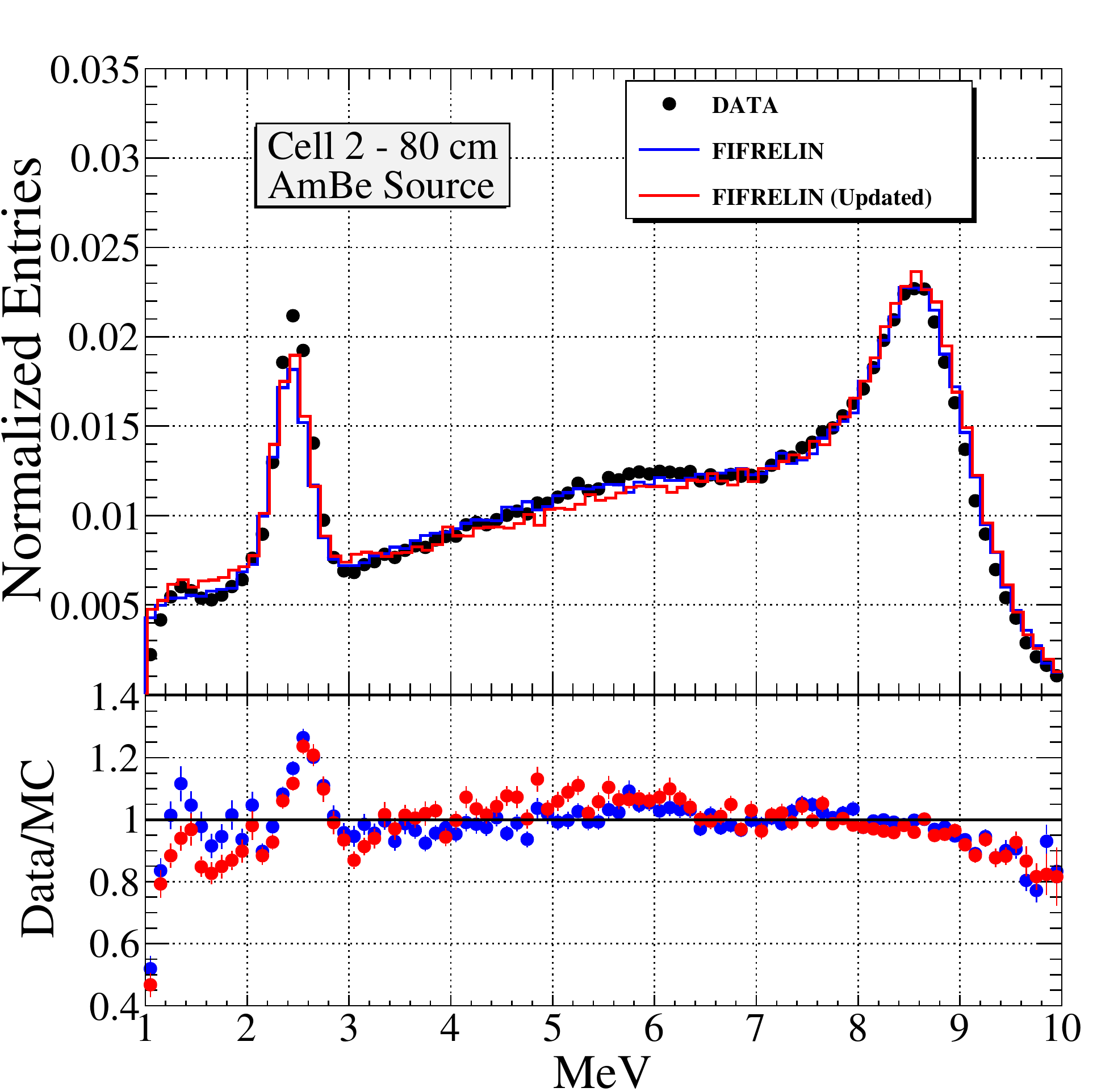}
			\centering
			\caption{}
			\label{fig: cell top}
		\end{subfigure}%
		
		\caption{(a) Reconstructed energy spectrum of the STEREO detector in coincidence with 4.4 MeV  prompt $\gamma$ ray from an AmBe source placed at the calibration position of cell 4 at 45 cm (center of the detector cell). (b) The same for cell 2 at 80 cm (10 cm from the top of the detector cell). The simulations with the updated FIFRELIN cascades (using RIPL-3 v.2020 and including EGAF transitions, angular correlations and X rays) is shown in red. The previous version of FIFRELIN cascades~\cite{Improved_stereo_Almazan2019}, which use RIPL-3 v.2015, are shown in blue. Experimental data are shown in black. See text for details. }
		
		\label{fig: cells results}
	\end{figure*}
	
	An identical analysis is applied on simulations and data, and a more detailed description can be found in~\cite{Allemandou_2018_The_STEREO_EXP}. The resulting delayed energy spectra are presented in Fig.~\ref{fig: cells results} and are compared with simulations using both the updated FIFRELIN version and the previous one from~\cite{Improved_stereo_Almazan2019}. The energy reconstruction corresponds to the whole STEREO detector. The 2.2 MeV peak from H(n,$\gamma$) and the $\sim$8 MeV from Gd(n,$\gamma$) are clearly visible. There is a very good agreement between both versions of FIFRELIN, as in particular positions inside the detector there seems to be a minor influence of the aforementioned updates in FIFRELIN.

	For a central position of the source (Fig.~\ref{fig: cell center}), most of the $\gamma$ rays of the Gd-cascade are contained in the detector and the Gd-peak is dominant. When approaching the border of the target volume (Fig.~\ref{fig: cell top}) a large fraction of the emitted $\gamma$ rays can escape the active volume, transferring the reconstructed events from the full energy peak to the lower energy tail. This change of shape of the neutron-capture spectrum is well described by both FIFRELIN simulations.

\begin{figure*}[ht!]
		\centering
		\begin{subfigure}[c]{0.47\textwidth}
			\includegraphics[width=\textwidth]{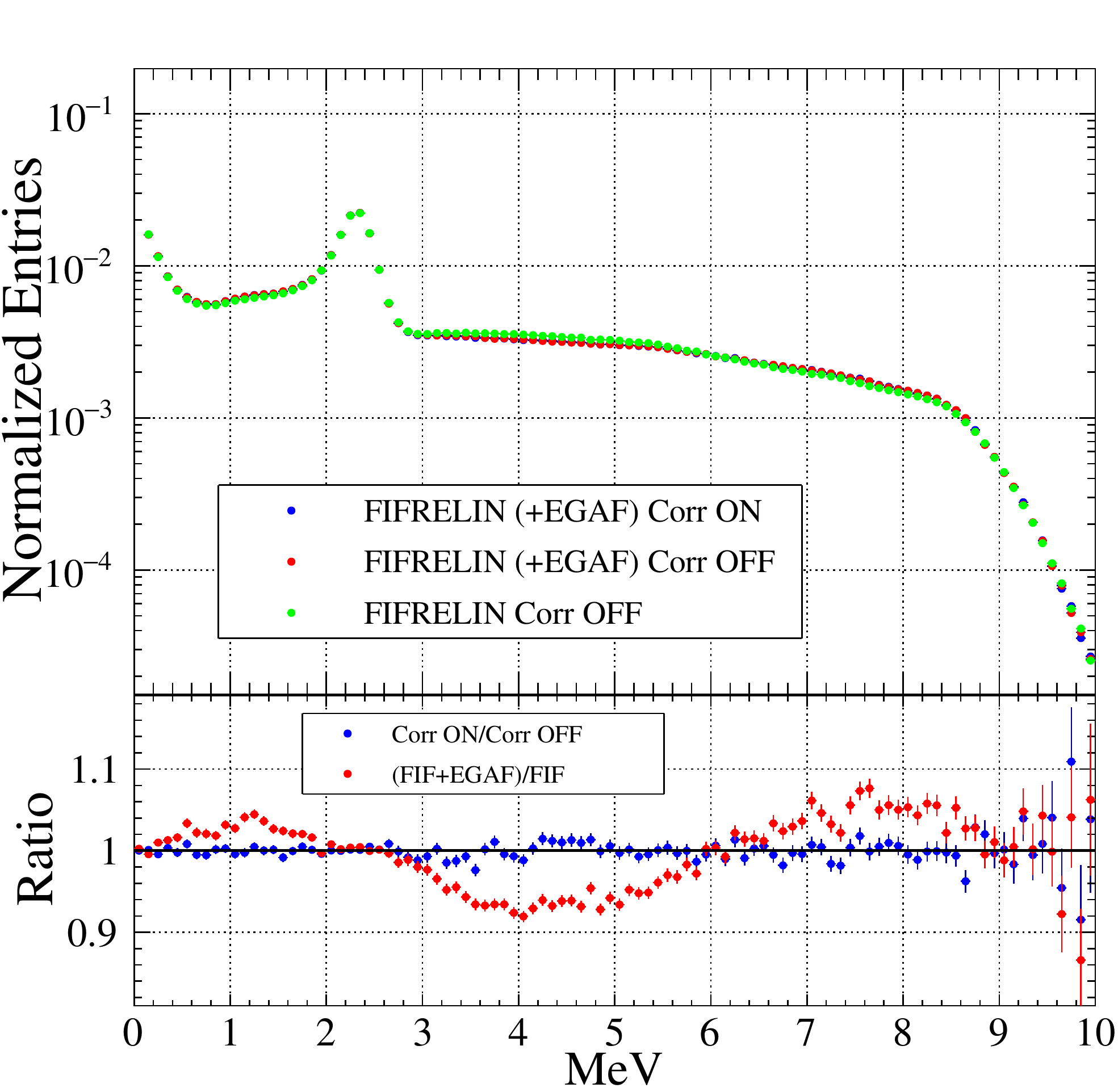}
			\centering
			\caption{}
			\label{fig: 555 corronoff}
		\end{subfigure}%
		\begin{subfigure}[c]{0.47\textwidth}
			\includegraphics[width=\textwidth]{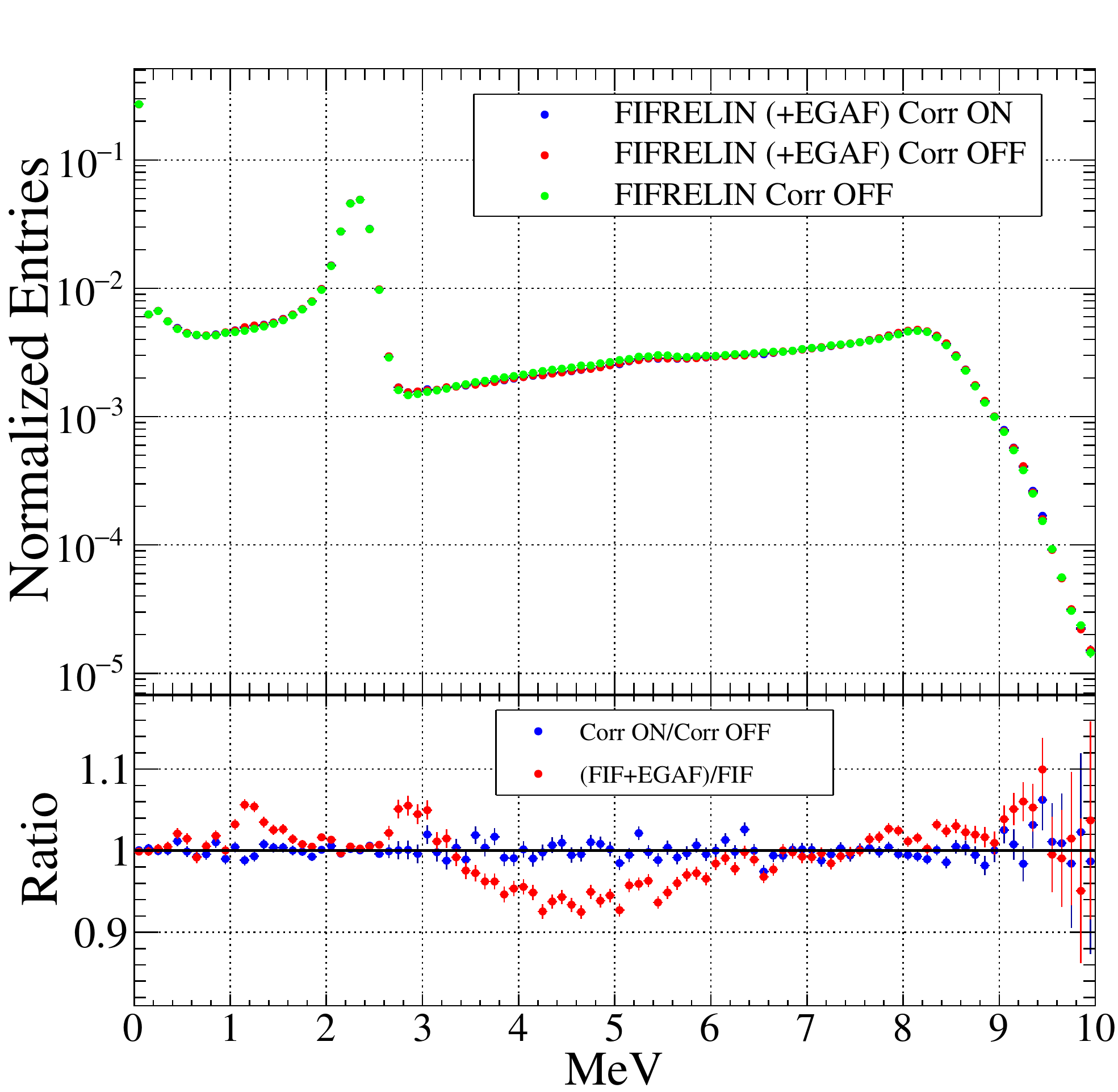}
			\centering
			\caption{}
			\label{fig: 55c corronoff}
		\end{subfigure}
		\caption{(a) Reconstructed energy spectrum of the STEREO detector obtained from a simulated neutron source placed close to the edge of Cell 1. "Corr ON" refers to enabled angular correlations of the $\gamma$ rays while Corr OFF refers to correlations being disabled. FIFRELIN cascades with and without the inclusion of EGAF transitions are also compared. The position of a neutron source is located at (5,5,5) mm from the top corner of Cell 1, which constitutes also the corner of the neutrino target of the detector. In (b) the same results are shown, but now with the neutron source located at (5,5) mm horizontally and at the center of the cell vertically.  See text for details. }
		
		\label{fig: cells results border}
	\end{figure*}

	However, the situation is not the same when neutron-captures are happening near the border of the detector cells, where there are no experimental data. In order to further check the impact of the new improvements in FIFRELIN beyond the calibration positions of the STEREO detector, simulations were also run near the border of the detector. A simulated neutron source is placed at various positions near the border of Cell 1.  
	A simulation on the border can provide insight on various changes on the shape of the spectrum of the energy deposited in the detector, as more $\gamma$ rays are prone to escape. The effect of the updated FIFRELIN cascades in the reconstructed spectrum near the border of the STEREO detector is demonstrated in Fig.~\ref{fig: cells results border}. 
	
	Two simulations were run in order to estimate the effect of EGAF primary transitions. In Fig.~\ref{fig: 555 corronoff}, the neutron source was placed at (5,5,5) mm with respect to the outer corner of Cell 1. There is an observable change in the shape of the spectrum when primary transitions from EGAF are included in FIFRELIN. This can be expected, as now the primary transitions are more intense, as shown in Fig.~\ref{fig: 156 egaf}, leading to a larger amount of $\gamma$ rays escaping the detector. It is important to note that in positions near the border of the detector there are no experimental data from STEREO. However, the very good agreement of the FIFRELIN spectrum with the evaluated data from ENSDF provides a strong argument for the use of the improved cascades. The same effect is demonstrated for a different position of the source in Fig.~\ref{fig: 55c corronoff}. The neutron source is now placed at (5,5) cm with respect to the corner but at the center of the cell in the z coordinate.
	
	Two simulations were also run in order to compare the effect of angular correlations. The results in this case show that the changes are not statistically significant. One explanation for this could be that the directions are averaged out in the detector leading to the same energy deposition inside the detector, with a negligible effect on the present spectra. However, since this effect is sensitive to the geometry of the experimental setup, other applications may benefit.

	\section{Discussion and future directions}
	
	New simulated cascades for the de-excitation of \isotope[156,158]{Gd} have been generated using the FIFRELIN code, to be used by the community for various applications. This code provides a refined description of the $\gamma$ cascades following neutron captures by gadolinium nuclei. Three main new features have been implemented for the simulations discussed here: 1) A complete set of measured primary $\gamma$ rays is built by merging the RIPL-3 and EGAF databases. 2) A full treatment of angular correlations is implemented. 3) The physics of IC and IPC processes is treated more accurately, including the secondary emission of X rays.   
	
	Recent improvements towards a more realistic de-excitation model of \isotope[156, 158]{Gd} can benefit the community in a wide range of applications in both low- and high- energy physics. The comparison with the STEREO data in Fig.~\ref{fig: cells results} show that both versions of FIFRELIN are able to describe well the reconstructed experimental spectrum. However, there are significant changes in the spectrum shape when the neutron capture is happening in positions close to the border of the cells. When compared with evaluated data, the de-excitation description using the new FIFRELIN cascades has been significantly improved especially for the higher energy part of the cascade, as seen in Fig.~\ref{fig: EGAF}. This constitutes a strong argument for the necessity of delivering an updated version of the previous FIFRELIN cascades to the community. A more precise description of the primary $\gamma$ rays is an improvement which can benefit applications and experimental setups that are sensitive to the higher energy part of the Gd spectra.
	
	The inclusion of $\gamma$-directional correlations is also an important aspect of the present work. Despite the negligible impact when applied for the STEREO detector, such an effect may be more pronounced in different experimental setups, making the angular correlations an essential aspect of the de-excitation process. Angular correlations are sensitive to the geometry of the detectors/setups, thus the necessity of their inclusion depending on each application can be very important. 
	
	The addition of X rays and the improvements on the electron conversion processes are also new features which improve the de-excitation description. The IC process is now treated more accurately, allowing for the emission of an X ray after an emission of a conversion electron. The improved modeling of the IC and IPC processes can be suitable in various experimental applications that rely on electron spectroscopy~\cite{MOUKADDAM2018180, Tee_2019_PhysRevC.100.034313, Chakma2020}.
	
	In conclusion, we make available ten millions of updated de-excitation cascades for the isotopes \isotope[156]{Gd} and \isotope[158]{Gd}~\cite{zenodo}, free of use for any other running and upcoming projects using neutron capture on gadolinium. The generalization of these $\gamma$-ray cascade predictions for other isotopes of interest is underway, notably in the context of cryo-detector calibration using neutron capture~\cite{Thulliez_2021}.

	\label{sec:discussion}

	%-----------------------------------------------------------------------------

	\section*{Acknowledgments}
	We acknowledge the financial support of the Cross-Disciplinary Program on Numerical Simulation of CEA, the French Alternative Energies and Atomic Energy Commission.

	\bibliographystyle{apsrev4-1}
	\bibliography{stereo}
\end{document}